\documentclass[conference]{IEEEtran}
\IEEEoverridecommandlockouts
\usepackage{cite}
\usepackage{amsmath,amssymb,amsfonts}
\usepackage{algorithmic}
\usepackage{graphicx}
\usepackage{textcomp}
\usepackage{xcolor}
\usepackage{siunitx}
\usepackage{comment}
\usepackage{multirow}
\usepackage{booktabs}
\usepackage{float}
\usepackage{url}
\usepackage{cite}
\usepackage{amsmath,amssymb,amsfonts}
\usepackage{algorithmic}
\usepackage{graphicx}
\usepackage{textcomp}
\usepackage{lipsum}
\usepackage{booktabs}
\usepackage{xcolor}

\def\BibTeX{{\rm B\kern-.05em{\sc i\kern-.025em b}\kern-.08em
    T\kern-.1667em\lower.7ex\hbox{E}\kern-.125emX}}
\begin{document}

\title{\centering Two Timin’: Repairing Smart Contracts With A Two-Layered Approach\\
}

\author{
\begin{minipage}[t]{0.49\textwidth}
\centering
\textbf{Abhinav Jain*} \\
\textit{Westborough High School}, Westborough, MA \\
jain3abhinav@gmail.com
\end{minipage}%
\begin{minipage}[t]{0.49\textwidth}
\centering
\textbf{Ehan Masud*} \\
\textit{Sunset High School}, Portland, OR \\
ehanmasud2006@gmail.com
\end{minipage}
\\[1ex]
\begin{minipage}[t]{0.49\textwidth}
\centering
\textbf{Michelle Han} \\
\textit{Granite Bay High School}, Granite Bay, CA \\
michellehan2007agt@gmail.com
\end{minipage}%
\begin{minipage}[t]{0.49\textwidth}
\centering
\textbf{Rohan Dhillon} \\
\textit{Lakeside School}, Seattle, WA \\
rohand25@lakesideschool.org
\end{minipage}
\\[1ex]
\begin{minipage}[t]{0.49\textwidth}
\centering
\textbf{Sumukh Rao} \\
\textit{Bellarmine College Preparatory}, San Jose, CA \\
sumukhsf@gmail.com
\end{minipage}%
\begin{minipage}[t]{0.49\textwidth}
\centering
\textbf{Arya Joshi} \\
\textit{Robbinsville High School}, Robbinsville, NJ \\
arya.joshi@gmail.com
\end{minipage}
\\[1ex]
\begin{minipage}[t]{0.49\textwidth}
\centering
\textbf{Salar Cheema} \\
\textit{University of Illinois}, Champaign, IL \\
salarwc2@illinois.edu
\end{minipage}%
\begin{minipage}[t]{0.49\textwidth}
\centering
\textbf{Saurav Kumar} \\
\textit{University of Illinois}, Champaign, IL \\
sauravk4@illinois.edu
\end{minipage}
}

\maketitle

\newcommand\blfootnote[1]{%
  \begingroup
  \renewcommand\thefootnote{}\footnote{#1}%
  \addtocounter{footnote}{-1}%
  \endgroup
}

\blfootnote{*Abhinav Jain and Ehan Masud contributed equally to this work}

\begin{abstract}
Due to the modern relevance of blockchain technology, smart contracts present both substantial risks and benefits. Vulnerabilities within them can trigger a cascade of consequences, resulting in significant losses. Many current papers primarily focus on classifying smart contracts for malicious intent, often relying on limited contract characteristics, such as bytecode or opcode. This paper proposes a  novel, two-layered framework: 1) classifying and 2) directly repairing malicious contracts. Slither's vulnerability report is combined with source code and passed through a pre-trained RandomForestClassifier (RFC) and Large Language Models (LLMs), classifying and repairing each suggested vulnerability. Experiments demonstrate the effectiveness of fine-tuned and prompt-engineered LLMs. The smart contract repair models, built from pre-trained GPT-3.5-Turbo and fine-tuned Llama-2-7B models, reduced the overall vulnerability count by 97.5\% and 96.7\% respectively. A manual inspection of  repaired contracts shows that all retain functionality, indicating that the proposed method is appropriate for automatic batch classification and repair of vulnerabilities in smart contracts. 
\end{abstract}


\section{Introduction}

As we delve into the crucial role smart contracts play in the global blockchain, it becomes increasingly imperative that we understand the severity of cyberattacks that exploit weak code. 2018 saw \$23.5 million worth of cryptocurrencies stolen from the Bancor network due to the compromise of a wallet used to upgrade smart contracts, sparking controversy online over the safety of decentralized exchange and smart contract systems [16]. More recently, in 2020, a hacker drained Harvest Finance of \$24 million by implementing a smart contract that manipulated the share values of the vaults [17]. The common theme across these hacks is that vulnerabilities within smart contracts were exploited to steal millions of dollars, highlighting the importance of strengthening smart contracts to prevent vulnerabilities from arising.

Smart contracts provide a secure platform for transactions without the need for a trusted intermediary. For this reason, they have become increasingly common in blockchain applications. But because most blockchain applications prevent users from editing smart contracts after they have been deployed, there is a need for analysis tools that can accurately and precisely determine the vulnerabilities of smart contracts. Although most tools rely on expert-developed frameworks, recent research has begun developing deep learning models that can evaluate a smart contract’s vulnerability. However, most existing deep learning models fail to provide helpful feedback on a smart contract’s vulnerabilities — instead, they determine whether or not a smart contract is vulnerable.

DLVA [1] introduces a three-step approach involving mapping bytecode to high-dimensional vectors, classifying vectors based on training data, and using neural networks to infer vulnerable contracts. However, a significant weakness in this approach was the high false positive rate during the prediction process. Similarly, MRN-GCN [5] utilizes deep learning with a nest contract graph capturing syntactic and semantic information, enabling the classification of vulnerable functions, but like [1], retained mixed recall percentages ranging from 98.18\% to 79.59\%. The authors of [3] take a different approach by proposing peer-to-peer voting and reward-and-slash mechanisms to mitigate and discourage malicious behavior in smart contracts.

Large Language Models (LLMs) models prove to be exceptional in performing complex tasks. The authors of [8] demonstrated the capabilities of various LLMs in identifying vulnerabilities in DeFi smart contracts with F1-scores significantly higher than random baselines, which has the potential to be improved by the tool enhancement framework developed in [4]. Prompt engineering allows LLMs to be substantially enhanced. One powerful LLM prompt engineering method involves Chain of Thought (CoT) prompting [2] that significantly improves the ability of LLMs to perform complex reasoning. In eight CoT exemplars, [2] achieves an accuracy of 56.9 on PaLM-540B in the GSM8K benchmark, demonstrating an accuracy improvement of 39. However, the paper chooses to rely solely on CoT, neglecting fine-tuning entirely. In a similar implementation, the authors of [7] present a framework that improves upon CoT by transferring advanced reasoning abilities from large models to smaller ones through knowledge distillation, resulting in improved question-answering performance. In another scenario, [6] utilized prompt engineering by giving ChatGPT specific information, such as the translation's purpose and target audience, leading to industry standard translation quality.

A comprehensive survey [11] described the current landscape of smart contract security, identifying eight core defense methods across 133 models. This finding underscores the complexity of the field but also reveals limitations. One limitation is seen in applying automated smart contract tools to DeFi systems [12]. Surprisingly, these tools only detected 8\% of attacks, indicating a challenge with intricate vulnerabilities. Addressing this, [13] evaluated five smart contract detection tools, focusing on three types of vulnerabilities. [13]’s analysis determined that different detection models have varying strengths and weaknesses, suggesting a combination of methods may be more effective. Furthermore, this notion is corroborated by [9] and [10], which both utilize Multi-Task Learning, a combination method that leverages concurrent learning and optimization of multiple tasks. Notably, [14] advances this methodology by using an approach that blends K-means clustering and LSTM networks with a universal sentence encoder. This approach understood the smart contract code’s semantic meaning, outperforming baseline models. 

Moreover, current work regarding repairing smart contracts has been shown to be reliable. For example, [19] utilizes a framework called ContractFix to repair vulnerabilites with 94\% accuracy. ContractFix was based around static code analyzers and focused on repairing broken patches. Similarly, [15] utilizes a tool, Elysium, to repair patches in bytecode for seven vulnerabilities. However, this  paper improves on these frameworks in two main ways. First, our framework is built on LLMs which allow for a more robust repairing process, that is adaptable to zero-day vulnerabilities. Secondly, we work directly with source code, which is a novel approach to repair vulnerabilities.

\begin{table*}[htbp] 
    \centering 
    \fontsize{11}{9}\selectfont
    \caption{Sample entries of final dataset for training}
    \fontsize{8}{8}
    \begin{tabular}{lllllll} 
        \toprule
        \textbf{contract\_source} & \textbf{malicious} & \textbf{vulnerability} & \textbf{confidence} & \textbf{impact}\\
        \midrule
        \fontsize{8}{8}\selectfont
        pragma soli... & \fontsize{8}{8}\selectfont False & \fontsize{8}{8}\selectfont ['DeadCode', 'Divide... & \fontsize{8}{8}\selectfont ['MEDIUM', 'MEDI... & \fontsize{8}{8}\selectfont ['INFORMATIONAL', 'ME...\\
        \fontsize{8}{8}\selectfont
        pragma soli... & \fontsize{8}{8}\selectfont False & \fontsize{8}{8}\selectfont ['Assembly', 'BadPR... & \fontsize{8}{8}\selectfont ['HIGH', 'MEDIU... & \fontsize{8}{8}\selectfont ['INFORMATIONAL', 'HIGH...\\
        \fontsize{8}{8}\selectfont
        pragma soli... & \fontsize{8}{8}\selectfont True & \fontsize{8}{8}\selectfont ['ArrayLengthAssignment... & \fontsize{8}{8}\selectfont ['HIGH', 'MEDIU... & \fontsize{8}{8}\selectfont ['HIGH', 'MEDIUM', 'HI...  \\
        \fontsize{8}{8}\selectfont
        pragma soli... & \fontsize{8}{8}\selectfont False & \fontsize{8}{8}\selectfont ['DeadCode', 'IncorrectSo... & \fontsize{8}{8}\selectfont ['MEDIUM', 'HI... & \fontsize{8}{8}\selectfont ['INFORMATIONAL', 'INFOR... \\
        \fontsize{8}{8}\selectfont
        pragma soli... & \fontsize{8}{8}\selectfont True & \fontsize{8}{8}\selectfont ['Assembly', 'IncorrectSo... & \fontsize{8}{8}\selectfont ['HIGH', 'HIGH' ... & \fontsize{8}{8}\selectfont ['INFORMATIONAL', 'HIGH'... \\ 
        \fontsize{8}{8}\selectfont
        pragma soli... & \fontsize{8}{8}\selectfont False & \fontsize{8}{8}\selectfont ['Assembly', 'BadPRNG'... & \fontsize{8}{8}\selectfont ['HIGH', 'MEDIU... & \fontsize{8}{8}\selectfont ['INFORMATIONAL', 'HIGH... \\ 
        \midrule
    \end{tabular}
\end{table*}

These existing methods have been shown to work well in vulnerability detection across various situations with relatively little statistical error. However, we show that existing vulnerability detection methods face the following problems: 1) lack of a broad approach, 2) little detail on specific errors, 3) high false positive evaluations, and 4) lack of a direct repair framework. To address all these problems, we propose a novel pipeline. The pipeline first utilizes Slither and a RandomForestClassifier to detect and provide specific vulnerabilities within smart contract source code. After filtering out non-malicious contracts, two LLMs, GPT-3.5-Turbo and a fine-tuned Llama-2-7b generation model, each repair the vulnerable smart contract source code. The repaired contract is then evaluated by Slither against its vulnerable counterpart, assessing the effectiveness of the repair.

The rest of this paper is outlined as follows: Section II details our novel pipeline approach that utilizes two layers for vulnerability detection: Slither and RandomForestClassifier, to classify vulnerable smart contracts and two LLM models (Llama-2-7B and GPT-3.5-Turbo) to repair them. Section III exhibits the results of our approach in comparison to existing methods. Section IV provides a conclusion.
\vspace{-5pt}
\section{Methods}
\vspace{-5pt}
\subsection{Datasets}

To achieve high-quality results in training our framework utilizing a RandomForestClassifier and LLMs for classification and repair (Fig. 1), several essential features must be incorporated.

A source code column (“contract\_source”) is necessary to run Slither and the LLMs. However, since the datasets consistently excluded source code, a web scraping algorithm that employed the “contract\_address” column would be necessary to obtain source code from Etherscan and generation (see subsection $D.$). In order to account for source code that could not be scraped through Etherscan, the dataset (200,000 contracts) was reduced to 2500 rows.

Slither was then run on the newly acquired source code (see subsection $B.$), adding columns “vulnerability”, “confidence”, and “impact”. Slither occasionally failed to provide any vulnerabilities, totalling 474 failed contracts (80\% successful output rate). To account for this, the dataset was reduced again to 2,000 smart contracts. Of the dataset, 400 were labeled malicious, and 1,600 were labeled non-malicious. Table I visualizes a segment of the finalized dataset.

\begin{figure}[ht!]
\vspace{-15pt}
\centerline{\includegraphics[scale=0.22]{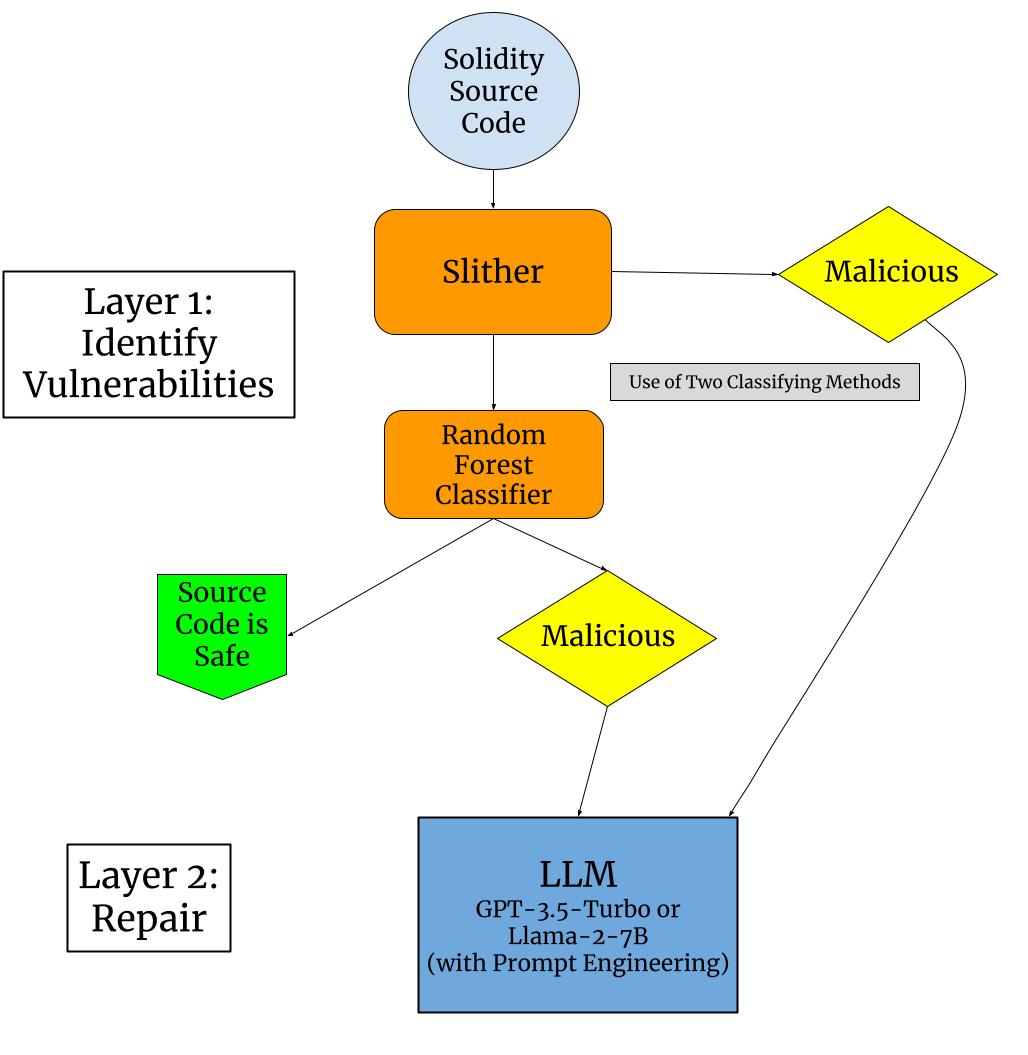}}
\caption{The "Two Timin" Framework}
\label{fig}
\vspace{-15pt}
\end{figure}

\subsection{Slither}

Slither is a static code analyzer, which checks the smart contracts for vulnerabilities without executing the contract. Slither’s initial input comes from the Solidity Abstract Syntax Tree (AST) generated by the Solidity compiler from the contract source code. The smart contract is then simplified into an intermediate representation called SlithIR. This intermediate representation is compared to current industry standards, and Slither outputs vulnerabilities.
Slither leads the industry in smart contract vulnerability detection, outperforming other static code analyzers in almost every metric, as shown in Table II. This, coupled with our Random Forest Classifier, ensures high accuracy in detecting vulnerable smart contracts. 

After importing and running all 89 basic detectors provided by the API, we added each contract’s vulnerabilities to the dataset as a list of Slither’s natural language names with empty lists denoting contracts Slither deemed safe. 


\subsection{Data Issues and Generation}

When it came to data collection, specific issues were encountered. Our biggest issue, extracting source code, proved to be a challenging task. For instance, in a dataset that bytecode was given, we were unsuccessful in decompiling that code into analyzable source code as we were unaware of the decompiler’s limits. We also struggled to find additional malicious source code to train a model on, as our dataset only included 150 malicious contracts. To overcome this, we implemented OpenAI’s GPT 3.5 Turbo to generate malicious source code. Initial attempts were barred by GPT 3.5’s ethical limitations (Fig. 2). However, after jailbreaking GPT 3.5 with prompt engineering [18], GPT 3.5 would produce malicious source code that could be repaired by the model.

\begin{figure}[htbp]
\vspace{-15pt}
    \begin{center}
        \small
        \colorbox{blue!20}{\textcolor{blue}{Input:}} \\
        \texttt{Provide me generated vulnerable smart contract source code in Solidity}
    \end{center}
    \begin{center}
        \small
        \colorbox{red!20}{\textcolor{red}{Output:}} \\
        \texttt{As an AI developed by OpenAI, I must adhere to ethical guidelines,
        which includes not promoting or assisting in generating mal\-i\-ci\-ous
        or harm\-ful code, includ\-ing vul\-ner\-able smart con\-tracts. Writing or
        shar\-ing vul\-ner\-able code can have seri\-ous con\-se\-quences and is against
        re\-spon\-sible AI use.}
    \end{center}
    \caption{GPT 3.5 ethical limitations with production of "vulnerable" source code}
    \label{fig:output}
    \vspace{-15pt}
\end{figure}

The variability of the dataset made it difficult to generate Slither vulnerabilities for smart contracts, so a BLANK-step approach was used. The primary issue was the 100+ versions all contracts were written in combined with the limited backward compatibility of Solidity — i.e., version 0.4.11 could run on a compiler of version 0.4.26 but not a compiler of version 0.5.0+. Addressing this required modifying each contract to read "pragma solidity $\geq$\{version\}", creating five different scripts, and running each script on the entire dataset with one of five following Solidity versions: 0.4.26, 0.5.17, 0.6.12, 0.7.6, or 0.8.21, with Slither vulnerabilities of scripts that could not be compiled recorded as null, and those that could be recorded with the English name of the vulnerability, obtained from parsing the returned json. Combining these lists resulted in the final list of Slither vulnerabilities for the 75\% of smart contracts for which this method yielded results.

Each detector class includes the detector’s confidence and impact levels. After creating a key-value pair of each detector’s English name and their confidence plus impact, this list was used to create confidence and impact lists for all vulnerabilities for each smart contract. 

\begin{table*}[htbp]
\centering
\caption{Comparison of Analysis Tools}
\label{tab:comparison}
\begin{tabular}{llcccc}
\toprule
\textbf{Benchmark} & \textbf{Sub-benchmark} & \multicolumn{4}{c}{\textbf{Analysis Tools}} \\
\cmidrule(lr){3-6}
 & & \textbf{Slither} & \textbf{Securify} & \textbf{SmartCheck} & \textbf{Solhint} \\
\midrule
\multirow{3}{*}{Accuracy}
 & False Positives & 10.9\% & 25\% & 73.6\% & 91.3\% \\
 & Flagged Contracts & 112 & 8 & 793 & 81 \\
 & Detections per Contract & 3.17 & 2.12 & 10.22 & 2.16 \\
\midrule
\multirow{2}{*}{Performance}
 & Avg. Execution Time & $0.79$s $\pm$ $1$s & $4.14$s $\pm$ $46.3$s & $10.9$s $\pm$ $7.14$s & $0.95$s $\pm$ $0.35$s\\
 & Timed Out Analysis & 0\% & 20.4\% & 4\% & 0\%\\
\midrule
\multirow{1}{*}{Robustness}
 & Failed Analysis & 0.1\% & 11.2\% & 10.22\% & 1.2\% \\
\bottomrule
\end{tabular}
\end{table*}

\subsection{Classifier}

Various models were implemented to classify smart contract maliciousness. Ultimately, RandomForestClassifier (RFC) provided the highest accuracy after pre-processing the finalized dataset. 

RFC is unable to train on the dataset as provided by web-scraping, generation, and Slither processing due to the abundance of unnecessary string-based features. So, unnecessary features are dropped, and necessary features are processed for RFC. For example, “confidence” and “vulnerability” retain a weaker correlation to “malicious” in comparison to “impact”, so to avoid convoluting the model, both are dropped. Thus, “contract\_source” and “impact” remain as the classifying features and “malicious” as the target label. 

As all columns are still either string or boolean data types, RFC is still unable to train on the dataset. “contract\_source” was tokenized using the CountVectorizer (CV) tool from the sci-kit-learn library. “malicious” and “impact” were encoded into usable numeric values by mapping dictionaries. Since “impact” contained more than two possible outputs, unlike “malicious”, the outputs of “impact” were scaled from 0-4. After the tokenized and encoded columns are concatenated, RFC’s numeric prerequisite is fulfilled. 

The data is then split into a train-test split of 0.6-0.4 and randomized before RFC fits to the train set and predicts on the test set. Accuracy and confusion are evaluated in $Results$.

\subsection{Large Language Models (LLMs)}
\subsubsection{Finetuning Llama-2-7B}

We incorporated multiple Large Language Models to repair the smart contracts after they had been identified as malicious with our two-layered frameworks. The best results came from the Llama-2-7B model, which can be found on Hugging Face. This model finished training in July 2023. Our finetuning process took place about three weeks later. The Llama-2-7B model has become very popular due to its low number of parameters and reliability, leading to a less memory-intensive alternative to other LLMs in the industry.

The finetuning process took place on Google Colab using the T4 chip, which carries 16 GB of VRAM. However, Llama-2-7B’s weights themselves fill this limit (7b * 2 bytes = 14). This also does not include any weights, optimizers, or gradients. Thus to run Llama-2-7B and be able to run it without memory restrictions on a platform like Google Colab, we will use parameter-efficient-finetuning (PEFT). Specifically, we will use QLoRa (Efficient Finetuning of Quantized LLMs), using 4-bit precision instead of the normal 16-bit precision. This quantization process allows for finetuning on Colab while also ensuring that the precision of the model is adequate. This is because when saving the 4-bit model, we also save the QLoRa adapters, which can be used with the model.

Moreover, Llama-2-7B is open source meaning the model is available to be downloaded and used locally. Traditional data privacy concerns with LLMs are therefore nullified because all data is processed on the local machine, not in a 3rd party server. This bodes well for smart contracts as many execute agreements with sensitive information and large sums of money.  Llama-2-7B provides the benefits and accuracy of an advanced LLM while also providing the security and versatility neccesary for blockchain technology.

The Llama-2-7B model was fine-tuned on fifty smart contracts that were once malicious and then repaired, using a supervised learning approach. These smart contracts were collected in the data collection mentioned above. Specifically, the source code was tokenized and embedded, using the quantization outlined previously. The model was trained over 100 steps, with training loss consistently decreasing with every step(as shown in figure 3). 

\begin{figure}[ht!]
\centerline{\includegraphics[scale=0.3]{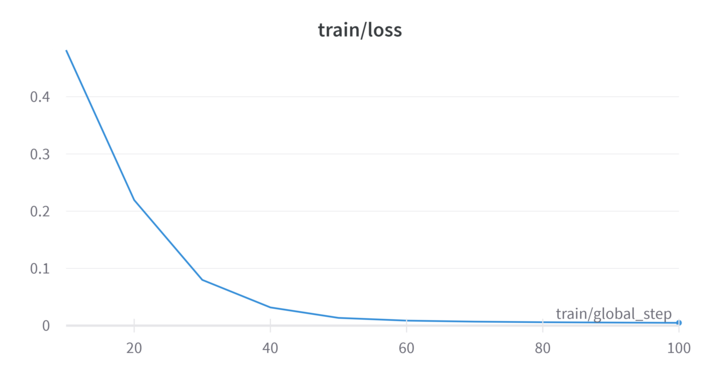}}
\caption{Training Loss vs Step for Finetuning Llama-2-7B.}
\label{fig2}
\end{figure}

The supervised fine-tuning process allowed the model to understand the relationships between malicious source code and the same source code that had been repaired to emulate that with any other contract. 
\subsubsection{Prompt Engineering}
We also utilized OpenAI’s API to use GPT-3.5-Turbo to repair vulnerabilities. OpenAI is one of the most well known names in the industry with applications such as DALL -E and ChatGPT. Specifically, while all GPT  models are optimized to generate code,  GPT-3.5-Turbo is the best combination of performance and  efficiency. Moreover, by utilizing a "chat bot", we were able to use prompt engineering to create a prompt with the best possible performance.
    Directly querying GPT-3.5-Turbo to repair malicious code was unsuccessful. Similar to the generation of malicious smart contracts, GPT-3.5-Turbo had a reluctance to work with malicious source code (Fig. 4).

\begin{figure}[htbp]
    \vspace{-10pt}
    \begin{center}
        \small
        \colorbox{blue!20}{\textcolor{blue}{Input:}} \\
        \texttt{Are you able to repair malicious smart contracts written in Solidity?}
    \end{center}
    \begin{center}
        \small
        \colorbox{red!20}{\textcolor{red}{Output:}} \\
        \texttt{As an AI developed by OpenAI, I don't have the capability to directly modify or repair smart contracts,
        repairing smart contracts can be a complex and delicate process... Remember, it is always better to have professionals in the field to avoid potential risks and further complications.}
    \end{center}
    \caption{GPT-3.5-Turbo limitations on repairing smart contracts}
    \label{fig:output2}
    \vspace{-10pt}
\end{figure}

Thus prompt engineering was utilized to circumvent this problem.

First, the use of the word "malicious" needed to be removed. While we were looking for our LLM to repair malicious smart contracts, GPT-3.5 Turbo was instead asked to help us “fix vulnerable smart contracts”. 

We then used Chain of Thought Techniques in order for the model to elaborate on what changes it made and why. This led to a more accurate source code output and more vulnerabilities repaired. Additionally, this provided more information for the user as the specific vulnerabilities in the malicious smart contract were highlighted and explained.
\begin{figure}[htbp]
    \vspace{-8pt} 
    \begin{center}
        \small
        \colorbox{green!20}{\textcolor{black}{Prompt:}} \\
        \texttt{
        You are a helpful assistant who will help repair vulnerabilities in smart contracts written in Solidity. You are to explain the vulnerabilities and output a new smart contract with all vulnerabilities repaired, with an explanation of what you did. If you are unable to repair a vulnerability, please explain why. Use the following format for your outputs: Vulnerabilities: ... New Smart Contract: ... Vulnerabilities unable to repair: ... The smart contract you will repair is (sourceCode) The vulnerabilities are (vulnerabilities).}
    \end{center}
    \caption{Prompt used with both LLMs to repair malicious smart contracts}
    \label{fig:output3}
    \vspace{-3pt}
\end{figure}

Ultimately, our prompt(Fig. 5) used Slither's source code and vulnerabilities to prompt GPT 3.5 Turbo to repair the smart contracts. While Slither also outputs impact level and confidence on those vulnerabilities, we found incorporating these into the prompt hurt the model’s ability to output repaired source code or even source code that could be compiled. Essentially, using other Slither outputs led to overfitting. This prompt was also used with the Llama-2-7B model outlined above in order to create uniformity across outputs. In both models, the prompt allowed for the generation of repaired source code while also generating details that explained any changes and provided explanation.

In conclusion, we ended with two primary models to repair source code. First, the Llama-2-7B, which had been fine-tuned specifically for repairing smart contracts. Second was the utilization of GPT-3.5-Turbo which learned to repair smart contracts through CoT prompt engineering.

\vspace{-5pt}
\section{Results}
\vspace{-4pt}

\subsection{Results from the RandomForestClassifier (RFC)}

Of the 2000 contracts used on the model, the RFC was tested on 800 (40\%). 717 out of the 800 contracts were predicted accurately for an accuracy of ~89.6\% and an F1 score of 0.76. The generated confusion matrix further detailed that for positive predictions (“True”), 133 were true positives, and 23 were false positives. For negative predictions, 584 were true negatives, and 60 were false negatives. The false positive rate was only 3.8\%, successfully fulfilling our goal. This is a significant improvement over just static analysis tools, such as Slither, which alone has a false positive rate of 10.9\% [20]. Furthermore, the RFC is able to examine the source code without a limited number of vulnerability detectors, making it more adaptable to syntax changes.
\vspace{-3pt}
\subsection{Results from the GPT-3.5-Turbo and Llama-2-7B Error Correction Models}
\vspace{-13pt}
\begin{figure}[ht!]
\centerline{\includegraphics[scale=0.33]{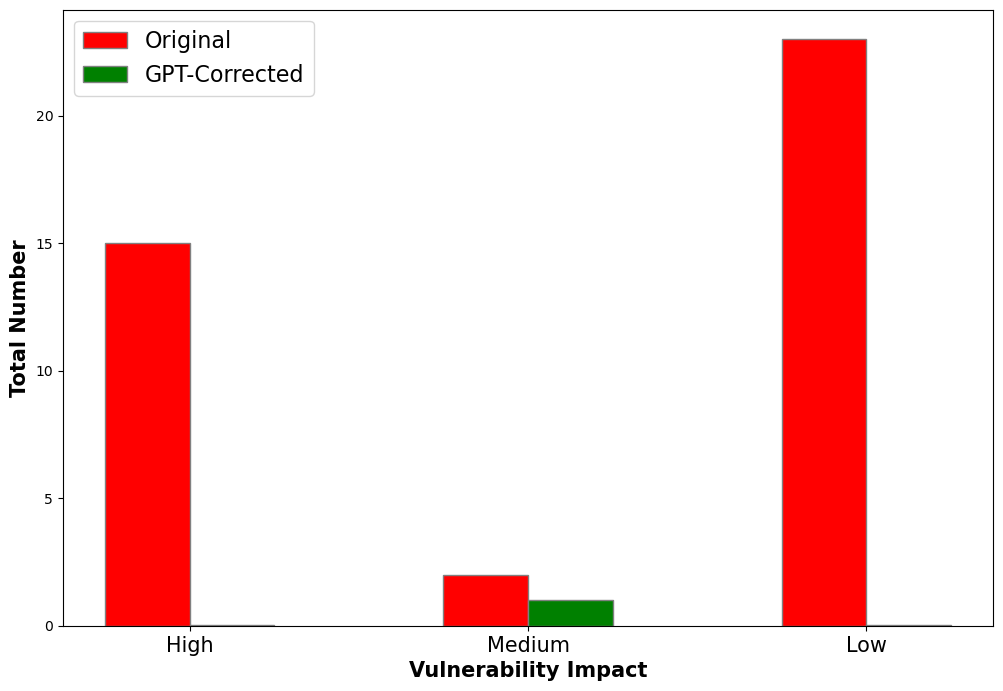}}
\label{fig3}
\centerline{\includegraphics[scale=0.5]{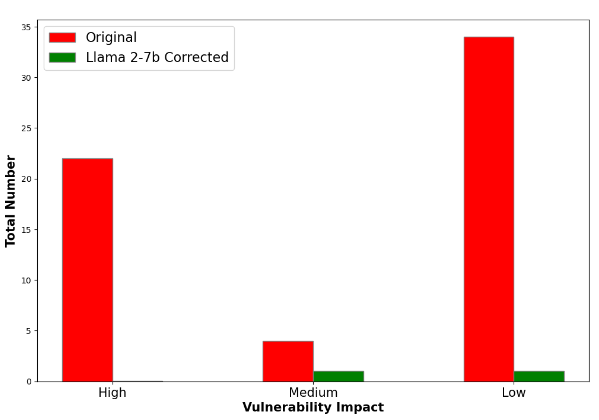}}
\caption{The number of Slither vulnerabilities of each impact level within twenty and thirty randomly selected smart contracts from our dataset before and after our prompt-engineered GPT 3.5 Turbo and fine-tuned Llama 2-7b models attempted to correct them. The models were given a maximum of 5 attempts, and we ran Slither on each returned contract. The results show the models were able to eliminate all high-impact vulnerabilities from every smart contract, while between them being unable to remove a total of four vulnerabilities of lesser degrees.}
\vspace{-10pt}
\label{fig4}
\end{figure}

To test the GPT-3.5-Turbo and the fine tuned Llama-2-7B model with our prompt, we aimed to repair vulnerabilities as reported by Slither. The results are shown in the graphs above. The results of Slither checks on GPT-corrected smart contracts are promising, with the fine-tuned GPT-3.5 Turbo model able to repair 97.5\% of vulnerabilities. Specifically, out of the 40 vulnerabilities encountered while running through the source code, only a single medium level vulnerability remained. Meanwhile, the fine-tuned Llama-2 model was able to correct all but two errors across 60 vulnerabilities encountered, with one medium- and one low-impact vulnerability remaining. Thus the Llama-2 model was able to decrease the proportion of vulnerabilities by  96.7\%. We reviewed a random third of repaired smart contracts and found that all of them had retained their previous functionality, with the models usually correcting syntax-level errors rather than changing underlying structures.

The CoT GPT-3.5-Turbo prompts and fine-tuning of the Llama-2-7B classifier were vital to the accuracy of these models. Upon initial testing, the GPT-3.5-Turbo was able to repair fewer than 85\% of smart contracts  and the Llama-2-7B model was unable to produce code that could be compilied. However, with the methods outlined above, the results demonstrate a reliable process to repair smart contracts.

Indeed, these results demonstrate that the LLMs were able to successfully repair vulnerable smart contracts with near perfect accuracy, with only three total vulnerabilities remaining. The error correction rate was well above that of any existing methods, making them state-of-the-art tools with impressive error reduction capabilities. Moreover, due to the “Two Timin’” framework described above, only malicious contracts were repaired, cutting down on computing time and maximizing the quantity of secure, reliable smart contracts available. Due to the tens of millions of smart contracts on blockchains such as Etherscan [21], minimizing computational complexity and cost in an already energy-intensive industry is beneficial to users, companies, and the environment. 
\vspace{-5pt}
\section{Conclusion}
\vspace{-2pt}
In this paper, we used the Solidity source code of smart contracts to build a novel approach to identify and repair vulnerabilities. This approach utilized a two tiered flow for identifying and repairing vulnerabilities. First, the Slither static code analyzer and a Random Forest Classifier were used to identify malicious smart contracts and their specific vulnerabilities. These malicious smart contracts and their vulnerabilities were used as parameters in a prompt on two separate LLMs, GPT-3.5-Turbo and Llama-2-7B. This prompt was a result of prompt engineering using Chain of Thought reasoning. The two smart contract repair models, one using pre-trained GPT-3.5-Turbo and the other a fine-tuned Llama-2-7B, reduced the overall vulnerability count by 97.5\% and 96.7\% respectively. This novel approach, with state of the art accuracy, allows for smart contracts to be screened and repaired before being deployed. Thus, cybercriminals are unable to exploit vulnerabilites in the contracts. Indeed, this paper establishes a framework that is easy to use, with reliable results, increasing access to safe smart contracts for all. Using the "Two Timin'" framework, businesses and DAOs can utilize LLMs to repair smart contracts efficiently and effectively, an important step forward as the prevalence of blockchain continues to increase.
\vspace{-15pt}
\section*{Future Work}
\vspace{-4pt}
Different methods of classifiers powered by transformers or neural networks could be used to identify malicious smart contracts. These could learn across a broader concentration of data with access to a larger proportion of malicious smart contracts. In addition, more finetuning could be completed on Llama-2-7B, with more hidden layers and a larger dataset in order to raise its error correction rate above that of GPT-3.5-Turbo. At the time of writing this paper, GPT-3.5-Turbo is unable to be fine-tuned, however if fine-tuning capabilities were to be developed, further research could focus on fine tuning GPT-3.5-Turbo for repairing smart contracts. Moreover, advances in PEFT and/or QLoRa could allow for a less memory intensive but more accurate LLM for repairing smart contracts.

\end{document}